\documentclass[aps,preprint]{revtex4}
\input epsf.tex
\def\DESepsf(#1 width #2){\epsfxsize=#2 \epsfbox{#1}}

\begin{document}
\preprint{\hbox{CTP-TAMU-19-00}}
\title{Yukawa Textures in Horava-Witten M-Theory} 

\author{R. Arnowitt and B. Dutta}

\address{ Center For Theoretical Physics, Department of Physics, Texas A$\&$M University,
College Station TX 77843-4242}
\date{June, 2000} 
\begin{abstract}Recent advances in 11 dimensional Horava-Witten M-theory based on non-standard
embeddings with torus fibered Calabi-Yau manifolds have allowed the construction of three
generation models with Wilson line breaking to the Standard Model gauge symmetry. Central to these
constructions is the existence of a set of 5-branes in the bulk. We examine within this framework
the general structure of the matter Yukawa couplings and show that M-theory offers an alternate
possible way of achieving the CKM and quark mass hierarchies without introducing undue fine tuning
or (as in conventional analysis) small parameters raised to high powers. A phenomenological example
is presented in accord with all CKM and quark mass data requiring mainly that the 5-branes cluster
near the second orbifold plane, and that the instanton charges of the physical orbifold plane
vanish. An explicit example of a three generation model with vanishing physical plane instanton 
charges based on a torus fibered
Calabi-Yau three fold with a del Pezzo base $dP_7$ and Wilson line breaking is constructed.
\end{abstract}
\maketitle

\section{Introduction} In spite of the remarkable success of the Standard Model in explaining
all low energy phenomena, there remains many unresolved questions concerning the structure
and nature of the Standard Model. Thus the standard Model contains a large number of
arbitrary parameters (mostly in the Yukawa sector) and has a pre-assigned particle spectra
and gauge structure. Low energy supersymmetry (e.g. the MSSM) inherits these same problems,
as well as the additional problem of how to spontaneously break supersymmetry. While
supergravity grand unified models (e.g. mSUGRA) allow for acceptable supersymmetry breaking
and hence the successful achievement of grand unification of the Standard Model gauge
coupling constants at $M_G\stackrel{\sim}{=}3\times 10^{16}$ GeV, the other questions
regarding the Standard Model remain unresolved.

In contrast, string theory models can in principle predict all the unexplained properties of
the Standard Model, and in fact is the only proposed theoretical structure that might be able
to make these predictions. In practice, however, the huge number of possible vacuum states,
and the incompleteness of the theory prevents actually realizing this. Thus, even from the
beginning of heterotic string theory, efforts were made to limit the theoretical choices by
imposing the phenomenological requirements of three generations of quarks and leptons with a
vacuum state  that could achieve the Standard Model gauge group at low energies \cite{y} and
attempts were made to further investigate the phenomenology of such systems \cite{ghmr}.

Similarly, the fact that grand unification does indeed occur at $M_G$ and not at the expected
Planck mass ($M_p=2.4\times 10^{18}$ GeV) was the phenomenological input leading to the
Horava-Witten M-theory \cite{hw1,hw2,w1}. In this model, 
one has the 11 dimensional orbifold structure
(to lowest order)
\begin{equation} M_4\times X\times S^1/Z_2
\label{eq1}
\end{equation} where $M_4$ is the Minkowski space, X is a (compact) Calabi-Yau 6 dimensional
space,  and $-\pi \rho\leq x^{11}\leq \pi\rho$. The gauge coupling constant at the GUT scale
$\alpha_G$ and Newtonian constatnt $G_N$ then obey 
\begin{equation}
\alpha_G={{(4 \pi \kappa^2)^{2/3}}\over{2 \mathcal{V}}};\,\,\,\, G_N={\kappa^2\over{16 \pi^2
\mathcal{V}\rho}}
\label{eq2}
\end{equation} where $\kappa^{2/9}$ is the 11 dimensional Planck scale and ${\mathcal{V}}$ is the
Calabi-Yau metric volume at the physical orbifold plane $x^{11}=0$. Setting
${\mathcal{V}}^{1/6}=1/M_G$ (so that grand unification occurs at the compactification scale)
and using $\alpha_G=1/24$ one finds
$\kappa^{2/9}\stackrel{\sim}{=}1/2 M_G$ and ${(\pi
\rho)}^{-1}\stackrel{\sim}{=}4.7\times 10^{15}$ GeV. Thus it is the 11 dimensional Planck
mass (which is the fundamental parameter) that sets the GUT scale, and  it is the largeness
of the orbifold size ($\pi \rho/\kappa^{2/9}\simeq 10$) that gives rise to the large four
dimensional Planck mass scaled by $(G_N)^{-1/2}$ (which is a derived parameter).

In M theory, the physical states live on a 10 dimensional manifold $M_4\times X$ at
$x^{11}$=0 and a hidden manifold exists at $x^{11}=\pi \rho$, each a priori having
$E_8$ gauge symmetry. In addition there can be a set of 5 branes in the bulk at points $x_n$
with $0<x_n<\pi\rho$, $n=1,...,N$ \cite{w1}, each with four dimensions spanning $M_4$ and wrapped
around a holomorphic curve in the Calabi-Yau space to preserve supersymmetry \cite{w1,bbs}.
 These act as additional hidden sectors
communicating with the physical sector by  gravity which lives in the bulk. Recently
there has been a great deal of analysis of this model, leading to the construction of three
generation models with low energy Standard Model gauge group
\cite{h1,bd,fmw1,fmw2,rd,low1,low2,gc,losw,lpt,low3,dlow1,dlow2,low4,o,lo,dopw,blo}.

The mathematics of these models is both rich and complex and while there are many possible
mathematical structures, presumably nature picks out only one. Phenomenological
considerations can therefore aid in deciding which are relevant. Thus an elegant possibility
is that the Calabi-Yau manifold be elliptically fibered, i.e. consist of a base B of two
complex dimensions fibered by elliptic curves (i.e. complex curves of genus one) and
possessing a global section $\sigma$. Such manifolds allow only a very restrictive set of
possible bases i.e. del Pezzo surfaces ($dP_r$, r=1...8), Hirzebruch surfaces ($F_m$) and
their blow-ups, and the Enriques surfaces ($\mathcal{E}$). (For a discussion of these manifolds 
see
\cite{dlow2,fr}.) However, it turns out that the first two  possibilities while having three
generation examples are simply connected manifolds and so do not allow the breaking of the
GUT gauge group to $SU(3)\times SU(2)\times U(1)$ by Wilson lines, while the last possibility
while allowing 
 for non-trivial Wilson lines is not consistent with three families. In order
to make contact with low energy physics, one may generalize the above by considering rather 
a set of torus fibered Calabi-Yau manifolds which allow for the same set of bases B but
possess two sections $\sigma$ and $\zeta$ related by a discrete
$Z_2$ symmetry \footnote{An alternate possibility based on $CP^N$ manifolds is considered in
ref.\cite{lo}.}. Such a manifold then can have  a Wilson line that can break
$SU(5)$ down to $SU(3)\times SU(2)\times U(1)$, and three family models of this type  have
been constructed in \cite{dopw}.

Torus fibered Calabi-Yau models thus represent an interesting class of models and in this
paper we investigate whether additional phenomenological constraints can be imposed. In
particular, the Cabibbo-Kobayashi-Maskawa (CKM) matrix, V$_{\rm CKM}$, and the quark masses
represent one of the major puzzles of the Standard Model, as it possesses a remarkable
hierarchy of parameters, e.g.
$m_u/m_t\simeq 10^{-5}$. The conventional field theoretic treatment of this question stems
from the original work of Georgi and Jarlskog \cite{gj} where one assumes a relatively simple
choice for the u and d quark Yukawa matrices $Y_U$,
$Y_D$ at $M_G$, and using renormalization group equations (RGEs) bring these down to the
electroweak scale where diagonalization gives rise to the V$_{\rm CKM}$ and the quark masses.
A general analysis assuming textures with five and six zeros in  $Y_U$ and $Y_D$ has been
given in \cite{rrr}. An example of such a model is given in Table 1 where $\lambda$ is the
Wolfenstein parameter.
 Note that one needs entries as small as $\lambda^{6}$ to appear in order to reproduce the
generation hierarchies observed at the electroweak scale,\footnote{Actually, the fits to the
low energy data of \cite{rrr} are no longer valid due to the improved accuracy in $V_{\rm cb}$
and $m_s$. An update of this will be given elsewhere
\cite{ad}.} and the origin of such small numbers remains obscure.\vspace{1cm}

{Table 1. Approximate Yukawa texture at $M_G$ for $Y_U$ and $Y_D$ where $\lambda$=0.2\cite{rrr}
 }\hrule{}\vspace{0.5cm}
$ Y_U=\left(\matrix{ 0 & \sqrt{2}\lambda^{6}  & 0 \cr
\sqrt{2}\lambda^{6} & \sqrt{3}\lambda^{4} & \lambda^{2} \cr 0  &
\lambda^{2}& 1 }\right); \,\,\,Y_D=\left(\matrix{ 0 &  2\lambda^{4}  & 0
\cr 2\lambda^{4} & 2\lambda^{3} & 0 \cr 0  & 0& 1 }\right).$\hrule{}\vspace{1cm}

The heterotic M-theory decribed above offers an alternate possibility for acquiring the
observed generational hierarchies. Here the Yukawa matrices are integrals over the Calabi-Yau
manifold, and while these integrals can not actually be performed, there is a priori no
expectation that the nonzero entries will not all be of the same size. However in addition to
$Y_{U,D}$ the theory possesses the Kahler metric $Z_{IJ}$ of the matter fields. This quantity
consists of two parts, an integral over the Calabi-Yau space $K_{IJ}$, with entries again
presumably all of 
$O(1)$ and a contribution from the 5-branes suppressed by factors $d_n=(1-z_n)$, $z_n=x_n/(\pi
\rho)$. In this paper we will assume $K_{IJ}$ contributes only to the first two generations,
while the 5-brane parts give rise to the third generation contributions. The Kahler metric
must be diagonalized and rescaled to the unit matrix  (so that the matter fields are
canonical) and it is this process that generates the Yukawa textures. Thus M-theory offers an
alternate way of thinking about the Yukawa textures. We find that in fact if the 5-branes
cluster near the hidden orbifold fixed point, i.e. $d_n\simeq 0.1$, the Yukawa textures can
naturally grow the necessary hierarchies.

Our paper is organized as follows: in Sec. 2 we review the basic results that  have been
obtained in heterotic M-theory for the torus fibered Calabi-Yau manifolds. In Sec. 3 we
generate the phenomenological Yukawa textures in the manner described above which are in
accord with all data. In the heterotic M-theory, the physical and the hidden orbifold planes carry instanton
numbers $\beta^{(0)}$ and  $\beta^{(M+1)}$, and the 5-branes magnetic charges $\beta^n$. In
order to get phenomenologically satisfactory Yukawa textures, it is necessary to consider
Calabi-Yau manifolds with 
$\beta^{(0)}=0$. This would not be possible for an elliptically fibered manifold \cite{lo} but
is feasible for torus fibered manifolds. In Sec. 4 we exihibit such a manifold for del Pezzo
base $dP_7$. In Sec. 5 we give concluding remarks.

\section{Heterotic M theory} In this section we summarize some of the recent results that
have been obtained in heterotic M-theory and discuss the general form of the Kahler metric
$Z_{IJ}$ for the physical fields $C^I$ on the 10 dimensional orbifold plane at $x^{11}$=0.
This will allow us to see what properties  might be imposed on the Kahler metric to obtain
phenomenologically acceptable Yukawa  matrices. With these conditions, an example of an
acceptable  set of Yukawa matrices is constructed in Sec. 3, and its phenomenology fully
examined there. In Sec. 4 we return to the M-theory analysis and exhibit explicitly an
example of a torus fibered Calabi-Yau manifold which satisfies at least part of the
requirements needed for the Yukawa matrices.

The Bose field content of 11-dimensional supergravity consist of two Yang-Mills potentials
and their field strengths, $A^{i}_R$, $F^{i}_{RS}$ ($i$=1,2) living on the orbifold planes at
$x^{11}$=0 and $x^{11}$=$\pi\rho$, the metric tensor,
$g_{IJ}$, the antisymmetric 3-form $C_{IJK}$ and its field strength $G_{IJKL}=24
\partial_{[I}C_{JKL]}$ living in the 11-dimensional bulk\footnote{We use $I$,$J$,$K$,... to
label the full 11 dimensional space 0,1,...9,11, while  $R$,$S$...=0,1,...,9 labels the 10
dimensional space on the orbifold planes. $A$,$B$...=4...9 label the (real) coordinates of
the Calabi-Yau space, and a,b,c and 
$\bar a$, $\bar b$, $\bar c$ =1,2,3 label their holomorphic and anti-holomorphic coordinates.
$\mu$, $\nu$, $\lambda$...=0,1,...,3 label the uncompactified Minkowski space. We normalize
our group generators such that $tr T^aT^b=2\delta_{ab}$.}. The $G_{IJKL}$ obeys equations of
motion $D_IG^{IJKL}=0$ and also possess a Bianchi identity \cite{low4}
\begin{eqnarray}\label{eq3}(dG)_{11RSTU}&=&4\sqrt 2\pi({\kappa\over
{4\pi}})^{2/3}[J^0\delta(x^{11})+J^{N+1}\delta(x^{11}-\pi \rho)\\\nonumber &+&{1\over
2}\Sigma^M_{n=1}J^{n}(\delta(x^{11}-x_n)+\delta(x^{11}+x_n))]_{RSTU}
\nonumber\end{eqnarray} where \begin{eqnarray}\label{eq4} J^{(0)}&=&-{1\over {16\pi^2}}(tr
F^{(1)}\wedge F^{(1)}- {1\over 2} tr R\wedge R)_{x^{11}=0}\\\nonumber J^{(N+1)}&=&-{1\over
{16\pi^2}}(tr F^{(2)}\wedge F^{(2)}- {1\over 2} tr R\wedge R)_{x^{11}=\pi\rho}
\nonumber\end{eqnarray}
$J^{(0)}$ and $J^{(N+1)}$ are the sources arising from the two orbifold planes while
$J^{(n)}$, $n$=1...$N$ are the sources from 5 the branes. ($R$ is the curvature tensor, the
wedge product acting on the first two indices). The Bianchi identity is even under the
orbifold $Z_2$ symmetry which requires that the 5 branes be placed symmetrically around
$x^{11}=0$.

Eq.(\ref{eq3}) may be thought of an expansion in powers of $\kappa^{2/3}$. Thus to lowest
order, one neglects the right hand side and hence $G^{(0)}_{RSTU}$ vanishes and the metric is
$g_{IJ}^{(0)}=(\eta_{\mu\nu}, g_{AB}, g_{11,11}=1$), where $g_{AB}$ is the Calabi-Yau metric.
One can then solve Eq.(\ref{eq3}) and
$D_IG^{IJKL}=0$ iteratively to get the first order correction \cite{low4}. We first note that if
we integrate Eq.(\ref{eq3}) over $x^{11}$ and any closed four dimensional surface in the
Calabi-Yau (i.e. a four cycle C$_4$), the left hand side vanishes since it is exact. Then 
\begin{equation}
\Sigma^{N+1}_{n=0}\int_{C_4} J^{(n)}=0
\label{eq5}
\end{equation} or using Eq.(\ref{eq4}) 
\begin{equation} -{1\over {16\pi^2}}\Sigma_i\int_{C_4}tr F^{(i)}\wedge F^{(i)}+ {1\over
{16\pi^2}}\int_{C_4}tr R\wedge R+\Sigma_{n=1}^N\int_{C_4}J^{(n)}=0
\label{eq6}\end{equation} If no 5-branes were present, Eq.(\ref{eq6}) reduces to the
condition leading to the ``Standard embedding" where one embeds the SU(3) spin connection
into the gauge group of the $x^{11}=0$ orbifold plane (See e.g.\cite{gsw}). More generally, each of the
terms in Eq.(\ref{eq6}) are integers and topological invariants. One defines the second Chern
class of each gauge group G on the orbifold planes as  
\begin{equation}c_2(V_i)=[{1\over {16\pi^2}} tr F^{(i)}\wedge F^{(i)}] 
\label{eq7}\end{equation} and the second Chern class of the Calabi-Yau tangent bundle as 
\begin{equation}c_2(TX)=[{1\over {16\pi^2}}\Sigma R\wedge R].
\label{eq8}\end{equation} Then cohomologically, Eq.(\ref{eq6}) implies
\begin{equation}c_2(V_1)+c_2(V_2)+[W]=c_2(TX)
\label{eq9}\end{equation} where $[W]=\Sigma^N_{n=1}[J^{(n)}]$ is the four form cohomology
class associated with the 5-branes.

In general, a Calabi-Yau manifold has a set of harmonic  (1,1) forms
$\omega_{ia\bar b}$ where $i=1...h^{(1,1)}$ and $h^{(1,1)}$ is the dimension of the
cohomology group $H^{(1,1)}(X)$. There is an associated set of independent 4-cycles $C_{4i}$,
and one can define the integer charges 
\begin{equation}\beta^{(n)}_i=\int_{C_{4i}}J^{(n)}
\label{eq10}\end{equation}$\beta^{(0)}_i$ and $\beta^{(N+1)}_i$ are the instanton charges on
the orbifold planes and $\beta^{(n)}_i$, n=1...N are the magnetic charges of the 5-branes
\cite{dml}. Eq.(\ref{eq5}) then implies that $\Sigma\beta^{(n)}_i=0$. We will see below that the
presence or absence of the $\beta^{(n)}_i$ plays an important role in attempting to construct
phenomenologically acceptable Yukawa matrices.

The existence of non-zero Yang-Mills fields with gauge group G on the orbifold plane implies
the breaking of the $E_8$ to a lower group. In general, $E_8$ will break into $G\times H$
where $H$ is the remaining symmetry of the physical theory. Thus if $G=SU(3)$ one has $H=E_6$
as in the standard embedding. For the physical orbifold plane at $x^{11}=0$ we will assume
here that\footnote{An alternate possibility not considered here is $G=SU(4)$ and then
$H=SO(10)$.} $G=SU(5)$ and hence $H=SU(5)$. A given representation then transforms as 
\begin{equation}\Sigma S_i\times R_i
\label{eq11}\end{equation} where $S_i$ is a representation of $G$ and $R_i$ of $H$. Thus
the 248 of $E_8$ decomposes under $SU(5)\times SU(5)$ as 
\begin{equation} 248=(24,1)\oplus (1,24)\oplus (10,5)\oplus (\bar{10},\bar{5})
\oplus (5,\bar{10})\oplus (\bar{5},10)
\label{eq12}\end{equation}

Consider now a physical field in representation $R$ of $H$ labeled by $C^{I}(R)$. Here $I$ is
the family index $I$=1,...,dim($H^1(X,S)$) where $H^1(X,S)$ is the cohomology group for the
associated representation $S$ in Eq.(\ref{eq11}). (We suppress the representation index of $R$
in
$C^I(R)$.) In general $H^1(X,S)$ has a set of basis functions $u^x_I(R)$ where
x=1,...,dim($S$) is the representation index for the representation $S$ associated with $R$
in Eq.(\ref{eq11}). One can also expand the Calabi-Yau metric in terms of a complete set of 
harmonic functions defined above i.e. (to zeroth order) 
\begin{equation} g_{a\bar b}=ia^i\omega_{ia\bar b};\,\, i=1,...,h^{1,1}
\label{eq13}\end{equation} where $a^i$ are moduli of the Calabi-Yau space. In terms of the
above quantities, one defines the metric \cite{low4}
\begin{equation} G_{IJ}(a^i;R)={1\over{vV}}\int_X\sqrt{g}g^{a\bar b}u_{Iax}(R)u^x_{J\bar b}(R)
\label{eq14}\end{equation} and the Yukawa couplings
\begin{equation}
\lambda_{IJK}(R_1,R_2,R_3)=\int_X\Omega\wedge u^x_{I}(R_1)\wedge  u^y_{J}(R_2)\wedge
u^z_{K}(R_3)f_{x,y,z}^{(R_1,R_2,R_3)}
\label{eq15}\end{equation} where $\Omega$ is the covariantly constatnt (3,0) form, f projects
out the gauge singlet parts, and ${\mathcal{V}}\equiv vV$ is the volume of the Calabi-Yau space
while
$v$ is the coordinate volume:
\begin{equation} V={1\over v}\int_Xd^6x\sqrt g;\,\,\,v=\int_Xd^6x
\label{eq16}\end{equation} In addition one defines the $S$, $T^i$ and 5-brane moduli by 
\begin{equation} Re(S)=V;\,\, Re T^i=Ra^i;\,\, Re Z_n=z_n
\label{eq17}\end{equation} where the modulus $R$ is the orbifold radius divided by $\rho$ and
$z_n=x_n/{\pi\rho}$.  $V$ can be expressed in terms of the $a^i$ moduli by $V(a)={1\over
6}d_{ijk}a^ia^ja^k$ where $d_{ijk}$ are the Calabi-Yau intersection numbers :
\begin{equation} d_{ijk}=\int_X\omega_i\wedge\omega_j\wedge\omega_k
\label{eq18}\end{equation} Following the techniques of \cite{w1}, the field equations and
Bianchi  identities in Eq.(\ref{eq3}) were solved in the presence of 5-branes to leading order
$O(\kappa^{2/3})$ \cite{low4} leading to an effective four dimensional Lagrangian at compactification
scale
$M_G$. We now state the results that were obtained. The gauge kinetic functions on the
orbifold planes are given by
\begin{eqnarray}\label{eq19} f^{(1)}&=&S+\epsilon
T^i(\beta^{(0)}_i+\Sigma^{N}_{n=1}(1-Z_n)^2\beta^{(n)}_i)\\\nonumber f^{(2)}&=&S+\epsilon
T^i(\beta^{(N+1)}_i+\Sigma^{N}_{n=1} Z_n^2\beta^{(n)}_i)
\nonumber\end{eqnarray} where 
\begin{equation}
\epsilon=\left({\kappa\over{4\pi}}\right)^{2/3}{{2\pi^2\rho}\over{{\mathcal{V}}^{2/3}}}
\label{eq20}\end{equation} The matter Kahler potential,
$K=Z_{IJ}\bar {C^I}C^J$, on the physical orbifold plane at
$x^{11}=0$ has the Kahler metric
\begin{equation}  Z_{IJ}=e^{-K_T/3}[G_{IJ}-{\epsilon\over{2V}}\tilde{\Gamma}^i_{IJ}
\Sigma^{N+1}_0(1-z_n)^2\beta^{(n)}_i]
\label{eq21}\end{equation} where 
\begin{equation} K_T=-ln[{1\over 6}d_{ijk}(T^i+\bar{T^i})(T^j+\bar{T^j})(T^k+\bar{T^k})]
\label{eq22}\end{equation}
\begin{equation}
\tilde{\Gamma}^i_{IJ}=\Gamma^i_{IJ}-(T^i+\bar{T^i})G_{IJ}-{2\over 3}
(T^i+\bar{T^i})(T^k+\bar{T^k})K_{Tkj}\Gamma^j_{IJ}
\label{eq23}\end{equation} and 
\begin{equation} K_{Tij}=\partial^2K_T/{\partial T_i\partial\bar{T^{j}}};\,\,\Gamma^i_{IJ}=
K^{ij}_T{{\partial G_{IJ}}\over {\partial T^j}}
\label{eq24}\end{equation} The Yukawa matrices are 
\begin{equation} Y_{IJK}=2\sqrt{2\pi\alpha_G}\lambda_{IJK}\simeq 1.02\lambda_{IJK}
\label{eq25}\end{equation} for $\alpha_G=1/24$. The Kahler metric on the distant orbifold
plane at
$x^{11}=\pi\rho$ is given by Eq.(\ref{eq21}) with $z_n\rightarrow(1-z_n)$.

\section{Phenomenological form of  The Yukawa matrices} We begin by discussing the general
structure of  Eqs.(\ref{eq19}-\ref{eq25}) to see under what circumstances they may yield
phenomenologicallly acceptable results. We see from Eq.(\ref{eq21}), that the Kahler metric
divides into parts : the first term proportional to $G_{IJ}$ of Eq.(\ref{eq14}), and the
second term scaled by the parameter 
$\epsilon$ of Eq.(\ref{eq20}) which measures the deformation of the manifold of
Eq.(\ref{eq1}). A priori, there is no reason to think that the integrals of Eq.(\ref{eq14})
are unduly suppressed, and one expects
$G_{IJ}=O(1)$. We note, however, that 
$\epsilon$ may not be too small, e.g. with the parameters of Sec.1, and assuming
${\mathcal{V}}^{1/6}\simeq 1/M_G$, one finds $\epsilon\simeq 0.93$. However, under certain
circumstances the second term in Eq.(\ref{eq21}) may indeed be a small correction to the
first. Recall that
$z_{N+1}=1$ and $z_0=0$ and so only the instanton charges
$\beta^{(0)}_i$ of the physical orbifold plane and the magnetic charges of the 5-branes,
$\beta^{(n)}_i$, $n$=1,...,$N$ enter in Eq.(\ref{eq21}). The 5-brane charges are shielded by
the factor $(1-z_n)^2$ and if the 5-branes were to cluster near the hidden orbifold plane i.e.
$z_n\simeq0.9$ one would get a strong suppression of the 5-brane terms. (Note the importance
of the quadratic structure, $(1-z_n)^2$, to achieve this without fine tuning.)

We conclude, therefore, that if $\beta^{(0)}_i$=0 and the 5-branes lie near the hidden
orbifold plane i.e.  $z_n\simeq0.9$, then the second term is naturally suppressed e.g. of
$O(10^{-2})$. In Sec 4. we will construct an example of a torus fibered Calabi-Yau that
allows $\beta^{(0)}_i=0$. In this section we will use the qualitative nature of the above
considerations to phenomenologically construct an example of Yukawa textures consistent with
the above that naturally explain the CKM and quark hierarchies without any significant fine
tuning. To do this we need also to assume that the metric $G_{IJ}$ (whose entries are assumed
to be $O(1)$) contributes in the u quark sector only to the first two generations, while the
second term of Eq.(\ref{eq21}) can contribute in general to all elements, but is the dominant
contribution to the third generation entries \footnote{In Sec. 4 we will also discuss what
properties of Calabi-Yau manifold might lead to $G^u_{33}$ and
$G^u_{32}$ being small.}. With the above set of assumptions, we now give an example in Table
2 of a phenomenologically acceptable set of Yukawa textures for the u and d quarks. In this
example we have assumed for simplicity that the Yukawa matrices are diagonal and that the
only one phase exists (in the third generation u quark entry). The quantity $f_T$ is given
from Eq.(\ref{eq21}) to be 
\begin{equation}f_T=Exp[-K_T/3]
\label{eq26}\end{equation} and we have assumed that the $Z^u$ and $Z^d$ have the maximum
number of zeros to make the model as predictive as possible. We also will assume here that the
$SU(5)$ symmetry holds at $M_G$ for the $Z$ matrices i.e. $Z^u$ is the Kahler metric for the
$u_R$, $u_L$ and $d_L$ quarks and $Z^d$ is the Kahler metric for the $d_R$
quarks\footnote{Since in torus fibered models, the group $H=SU(5)$ is broken to 
$SU(3)\times SU(2)\times U(1)$ by a Wilson line \cite{dopw}, it is not necessary for the $SU(5)$
symmetry to still hold in the Yukawa sector \cite{gsw}. We assume that $SU(5)$ symmetry still
holds to consider the simplest and most predictive model.}.\vspace{1cm}

{Table 2. Kahler matrices $Z_{IJ}$ and Yukawa matrices $Y^u$, $Y^d$ for u and d quarks for
tan$\beta$=3. The  parameter d is 0.1.}\hrule{}\vspace{0.5cm}

\begin{equation} Z^u=f_T\left(\matrix{ 1  & 0.345  & 0 \cr 0.345 & 0.132 & 0.639 d^2
\cr 0  & 0.639 d^2& 0.333 d^2 }\right); \,\,\,Z^d= 
f_T\left(\matrix{ 1 & 
0.821  & 0 \cr 0.821 & 0.887 & 0 \cr 0  & 0& 0.276  }\right).
\label{eq27}\end{equation}

\begin{eqnarray}{\rm diag}Y^u&=&{(0.0765,\, 0.536,\, 0.585 Exp[
\pi i/2])};\\\nonumber
                 {\rm diag}Y^d&=&{(0.849,\, 0.11,\, 1.3)}.  
\label{eq28}\end{eqnarray}\vspace{0.1cm}\hrule{}\vspace{1cm}

We first note two scale invariances of the Yukawa matrices. First, since the normalization of
the harmonic (1,1) forms $\omega_{ia\bar b}$ is arbitrary, one might rescale them by
$\omega_{ia\bar b}\rightarrow \mu\omega_{ia\bar b}$. From Eq.(\ref{eq13}) this implies $
a_i\rightarrow a_i/\mu$ and by Eq.(\ref{eq18}) then 
$ d_{ijk}\rightarrow \mu^3 d_{ijk}$. Thus $K_T$ and also $Re(S)=V$ are invariant while
$Re(T_i)\rightarrow (Re(T_i))/\mu$. From Eq.(\ref{eq25}) one sees $Y_{IJK}$ are invariant.
Similarly one may rescale the (1,0) $u_{Ia}$ forms by 
$u_{Ia}\rightarrow\lambda u_{Ia}$. This implies 
$G_{IJ}\rightarrow|\lambda|^2G_{IJ}$ and hence 
$Z_{IJ}\rightarrow|\lambda|^2Z_{IJ}$ while by Eq.(\ref{eq15})
$Y_{IJK}\rightarrow\lambda^3Y_{IJK}$. However to reduce the kinetic energy terms of the quark
fields to canonical form requires then a field rescaling of 
$C^I\rightarrow C^I/\lambda$, and hence the Yukawa interactions 
(i.e. Eq.(\ref{eq33}) below) are again scale invariant. In
Table 2, we have used the $u_{Ia}$ scale freedom to make the
$G_{IJ}=O(1)$ and in fact set $G_{11}=1$.

We proceed now as follows: we diagonalize and rescale $Z^{u,d}$ matrices at
$M_G$ to the unit matrix by a unitary and scaling transformation on the quark fields 
\begin{equation} C^I_{u,d}={1\over {\sqrt{f_T}}}(U^{(u,d} S^{(u,d})_{IJ}
{C_{u,d}^{J}}^{\prime}
\label{eq29}\end{equation} where $U^{\dag}=U^{-1}$ diagonalizes $Z$ and 
\begin{equation} {\rm diag} S= (\lambda_1^{-1/2},\, \lambda_2^{-1/2},\,\lambda_3^{-1/2}).
\label{eq30}\end{equation} Here $\lambda_i^{(u,d)}$ are the eigenvalues of the  $(u, d)$
contributions to
$Z_{IJ}/f_T$. In addition the Higgs fields have a contribution to $Z_{IJ}$ of the form 
\begin{equation} f_T G_{H_{1,2}}{\bar {H}}_{1,2}H_{1,2}
\label{eq31}\end{equation} and we make the corresponding rescaling of $H_{1,2}$:
\begin{equation} H_{1,2}={1\over {\sqrt{f_T G_{H_{1,2}}}}}H'_{1,2}
\label{eq32}\end{equation} The Yukawa contribution to the superpotential is \cite{low4}
\begin{equation} W_Y=e^{{1\over 2}K_m}{1\over 3}Y_{IJK}C^IC^JC^K
\label{eq33}\end{equation} where $K_m$ is the moduli contribution to the Kahler potential,
\begin{equation} K_m=-ln(S+\bar S)+K_T
\label{eq34}\end{equation} and hence by (\ref{eq17}) and (\ref{eq22})
\begin{equation} K_m=-ln(2 V)-ln({8}R^3V)
\label{eq35}\end{equation} In terms of the canonically normalized field variables, then, the
u and d quark contributions to $W_Y$ are 
\begin{equation} W^{(u)}={1\over{8{\sqrt 2}}}{1\over {R^3 V^{3/2}}}{1\over {\sqrt
G_{H_2}}}u'_L\lambda^{(u)}u'_RH'_2
\label{eq36}\end{equation} and
\begin{equation} W^{(d)}={1\over{8{\sqrt 2}}}{1\over {R^3 V^{3/2}}}{1\over {\sqrt
G_{H_1}}}d'_L\lambda^{(d)}d'_RH'_1
\label{eq37}\end{equation} where $\lambda^{(u,d)}$ are given by
\begin{equation}
\lambda^{(u)}_{IJ}=(S^{(u)}{\tilde U^{(u)}}Y^{(u)}U^{(u)}S^{(u)})_{IJ}
\label{eq38}\end{equation}
\begin{equation}
\lambda^{(d)}_{IJ}=(S^{(u)}{\tilde U^{(u)}}Y^{(d)}U^{(d)}S^{(d)})_{IJ}
\label{eq39}\end{equation} and $\sim$ means transpose. The $\lambda^{(u,d)}_{IJ}$ play the role of the Yukawa
textures in the  phenomenological analyses of e.g. Table 1\cite{rrr}. Note, however, that
$\lambda^{(d)}$ is not symmetric, and so the M-theory textures are uniquely different from
what has previously been considered.

Eq.(\ref{eq36}) and (\ref{eq37}) hold of course at $M_G$. One then uses the supersymmetry
renormalization group equations (RGEs) to evaluate the Yukawa couplings at low energy. We use
here the one loop Yukawa RGEs and two loop gauge RGEs\cite{bbo} to  generate the Yukawa
couplings at the electroweak scale which we take to be $m_t$\footnote{Note that in M-theory,
even though $H=SU(5)$ is broken to $SU(3)\times SU(2)\times U(1)$ at $M_G$ by a Wilson line,
the gauge coupling constants (which enter in the Yukawa RGEs) still unify according to
$SU(5)$\cite{gsw}.}. Below $m_t$ we assume the Standard Model holds, and include in our
calculations the QCD corrections (which are quite significant). Diagonalization of the Yukawa
couplings then allows one to generate the low energy quark masses and CKM matrix elements.
These are given in Table 3 for the model of Table 2 where we've set $G_{H_{1,2}}=1$. (We use
the QCD correction factors: $\eta_c=2$, $\eta_u=2.5=\eta_d$, $\eta_b=1.6$ and $\eta_s=2.5$.) 
\vspace{1cm}

{ Table 3. Quark and CKM matrix elements obtained from the model of Table 2. Masses are in
GeV. Experimental values are from \cite{pdg} unless otherwise noted.}\hrule{}
\begin{center} \begin{tabular}{|c|c|c|}  
 \hline Quantity&Theoretical Value&Experimental Value\\\hline
$m_t$(pole)&170.5& 175$\pm$ 5\\
$m_c$($m_c$)&1.36&1.1-1.4\cr
$m_u$(1 GeV)& 0.0032&0.002-0.008\\
$m_b$($m_b$)& 4.13&4.1-4.5\cr
$m_s$(1 GeV)& 0.110&0.093-0.125\cite{sa}\\
$m_d$(1 GeV)& 0.0055&0.005-0.015\\
$V_{us}$&0.22&0.217-0.224\\
$V_{cb}$&0.036&0.0381$\pm$0.0021\cite{ss}\\
$V_{ub}$&0.0018&0.0018-0.0045\\
$V_{td}$&0.006&0.004-0.013\cr sin$2\beta$&0.31&\\ sin$\gamma$&0.97&\\
\hline\end{tabular}\end{center}
\hrule{}\vspace{1cm}

The agreement between the theoretical values and current experimental data is very good. In
particular, the quark mass hierarchies are reproduced\footnote{We note that our results are
in close agreement with the phenomenological analysis of the light quark masses of \cite{hl}:
$m_u/m_d=0.553\pm0.043$ and
$m_s/m_d=18.9\pm 0.8$} even though no very small parameter enters in Table 2, unlike Table 1.
Though one needs a computer calculation to get precise values, one can qualitatively
understand these successes in an analytic fashion. To simplify the analysis, we use an
approximate form of $Z^u$ and $Z^d$ given in Table 4. The eigenvalues of
$Z^u/f_T$ of the first two generations can be computed approximately by neglecting $O(d^2)$
terms:
\begin{equation}
\lambda^2-{9\over 8}\lambda+{1\over 8}-{1\over 9}=0
\label{eq40}\end{equation} yielding roots $\lambda^{(u)}_1=1.13$, $\lambda^{(u)}_2=0.0123$.
(Using the accurate matrices of Table 2 one finds $\lambda^{(u)}_1=1.12$,
$\lambda^{(u)}_2=0.0144$.)  We see that $\lambda^{(u)}_2$ is a factor of 100 smaller than
$\lambda^{(u)}_1$ (i.e. a factor of 100 smaller than one might have expected), which exhibits
the fact that it is possible to generate a significant hierarchy without introducing any
parameter of smallness or overt fine tuning. From Eqs.(\ref{eq30}) and (\ref{eq38})
one sees that the factor
${1/\lambda_i}$ are major ingredients in generating the quark mass hierarchies, and thus it
is the above suppression of ${\lambda^{(u)}_2}$ that generates a large $m_c/m_u$. Turning
to the third eigenvalue of $Z^u/f_T$, one might expect (from Table 4) that
$\lambda^{(u)}_3\simeq (1/3)d^2\simeq0.0033$. However, this is not the case. To determine
$\lambda^{(u)}_3$ we use the full secular equation\begin{equation} ({1\over
3}d^2-\lambda)[\lambda^2-{9\over 8}\lambda+{1\over 72}]-{4\over 9}(1-\lambda)d^4=0
\label{eq41}\end{equation} If one could neglect the last $O(d^4)$ term, then indeed 
$\lambda^{(u)}_3\simeq1/3d^2$. However writing Eq.(\ref{eq37}) as 
\begin{equation}
\lambda[({1\over 3}d^2-\lambda)(\lambda-{9\over 8})-{1\over 72}+{4\over 9}d^4]+[{d^2\over
3}{1\over 72}-{4\over 9}d^4]=0
\label{eq42}\end{equation} one sees that because of the (1/72) factor (which caused the
suppression of 
$\lambda^{(u)}_2$) the last bracket is actually $O(d^6)$. Hence neglecting 
$\lambda$ in the first bracket (since $\lambda^{(u)}_3$ is very small) one finds
\begin{equation}
\lambda^{(u)}_3\simeq[{d^2\over 3}({1\over 72})-{4\over 9}d^4]/[{1\over 72}+{3\over
8}d^2-{4\over 9}d^4]=1.053 d^4
\label{eq43}\end{equation} Thus the mechanism that suppresses $\lambda^{(u)}_2$ propagates to
the third generation further suppressing $\lambda^{(u)}_3$, again without introducing any
parameters of smallness. (Actually the accurate matrices of Table 2 gives a 
$\lambda^{(u)}_3$ about three times smaller than the estimate of Eq.(\ref{eq43}), i.e.
 $\lambda^{(u)}_3=0.344 d^4$.) 
\vspace{1cm}

{ Table 4. Approximate $Z^{u,d}$ matrices for analytic analysis. (Compare with Table
2.)}\hrule{}\vspace{0.2cm}
\begin{equation}Z^u=f_T\left(\matrix{ 1  & 1/3  & 0 \cr 1/3 & 1/8 & (2/3) d^2 \cr
 0  &
(2/3) d^2& (1/3) d^2 }\right); \,\,\,Z^d= f_T\left(\matrix{ 1 &  4/5  & 0
\cr 4/5 & 9/10 & 0 \cr 0  & 0& 1/4 }\right).
\label{eq44}\end{equation}\vspace{0.2cm}\hrule{}\vspace{1cm}

 The above analysis also allows us to estimate the quark mass ratios. Thus when
 $H_{1,2}'$ grow VEVs,
\begin{equation} <H'_1>={v\over {\sqrt 2}}cos\beta;\, <H'_2>={v\over {\sqrt
2}}sin\beta;\,v=246{\rm GeV}
\label{eq45}\end{equation} one reads off from Eqs.(\ref{eq36}-\ref{eq39}), the u and d quark
masses to be:
\begin{equation} m_i^{(u)}={1\over{8{\sqrt 2}}}{1\over {R^3 V^{3/2}}}{1\over {\sqrt
G_{H_2}}}{\mu_i^{(u)}\over\lambda^{(u)}_i}{v\over\sqrt{2}}sin\beta
\label{eq46}\end{equation} and
\begin{equation} m_i^{(d)}={1\over{8{\sqrt 2}}}{1\over {R^3 V^{3/2}}}{1\over {\sqrt
G_{H_1}}}{\mu_i^{(d)}\over{\sqrt{\lambda^{(u)}_i\lambda^{(d)}_i}}}{v\over\sqrt{2}}cos\beta
\label{eq47}\end{equation} where ${\mu_i^{(u,d)}/\lambda^{(u,d)}_i}$ are the eigenvalues of
$\lambda^{(u,d)}_{IJ}$ at the electroweak scale. In writing Eqs.(\ref{eq46}-
\ref{eq47}) we have factored out the $1/\lambda^{(u,d)}_i$ factors that appear in $S^{(u,d)}$
 and so we expect $\mu_i^{(u,d)}=O(1)$. Thus $m^{(u)}_i$ scales with 
 $1/\lambda^{(u)}_i$, and the above suppression mechanism of the 
 $\lambda^{(u)}_i$ gives the relations
 \begin{equation}
 m_t\simeq100 m_c\simeq (100)^2m_u
 \label{eq48}\end{equation} in qualitative agreement with the experiment. The ratio $m_b/m_t$
is 
 \begin{equation}
 {m_b\over {m_t}}={\sqrt{G_{H_2}\over G_{H_1}}}
 {\mu^{(d)}_{3}\over \mu^{(u)}_{3}}
 {\sqrt{\lambda^{(u)}_{3}\over \lambda^{(d)}_{3}}}{1\over tan\beta}
 \label{eq49}\end{equation} From Table 4 one has $\lambda^{(d)}_{3}=0.25$, and using the
accurate values for
$\lambda^{(u)}_{3}$, one finds ($m_t$=175 GeV, $tan\beta$=3)
\begin{equation} m_b\simeq {\sqrt{G_{H_2}\over G_{H_1}}}{\mu^{(d)}_{3}\over
\mu^{(u)}_{3}}{1} {\rm GeV}
\label{eq50}\end{equation} giving the correct scale for $m_b$. One sees then that the reason
the $b$-quark is so much lighter than the $t$ quark is due to the fact 
$\lambda^{(u)}_3$ is anomalously reduced, but $\lambda^{(d)}_3$ is not, and that
$m_b\sim 1/{\sqrt{\lambda^{(u)}_3/\lambda^{(d)}_3}}$. The hierarchies of all the
$d$-quark masses then follow naturally using Table 2. Thus the down quark mass ratios are 
\begin{equation} {m^{(d)}_{i}\over m^{(d)}_{j}}={\mu^{(d)}_{i}\over
\mu^{(u)}_{j}}\left[{{\lambda_{j}^{(d)}\lambda_{j}^{(u)}}\over
{\lambda_{i}^{(d)}\lambda_{i}^{(u)}}}\right]^{1/2}
\label{eqa}\end{equation}
To obtain a qualitative understanding of these, we again use the approximation of Table 4
where one finds $\lambda^{(d)}_1\stackrel{\sim}{=}1.75$,
$\lambda^{(d)}_2\stackrel{\sim}{=}0.15$, $\lambda^{(d)}_3\stackrel{\sim}{=}0.25$.
Eq.(\ref{eqa}) then gives (using the accurate value $\lambda^{(u)}_3=3.44\times 10^{-5}$)
\begin{equation}{m_s\over m_d}\stackrel{\sim}{=}{\mu^{(d)}_{2}\over
\mu^{(d)}_{1}}33;\,\,{m_b\over m_s}\stackrel{\sim}{=}{\mu^{(d)}_{3}\over
\mu^{(d)}_{2}}15\label{eqb}\end{equation}
 in accord with the experimental values of $m_s/m_d\simeq20$,
$m_b/m_s\simeq40$, since the factors $\mu^{(d)}_{i}$ are $O(1)$.
 
 To summarize the above: the hierarchy of the up and down quark masses, and the fact that
 the $d_i$ quarks are significantly lighter than their $u_i$ quark counterparts
 (except $m_d$) has at its core the suppression of $\lambda^{(u)}_2$ which then
 generates a further suppression of $\lambda^{(u)}_3$, and this can occur in
 the above M-theory model without any very small parameters or extreme fine tuning
 entering.
 Thus the qualitative predictions of (\ref{eq48}), (\ref{eq50}) and (\ref{eqb}) follow fairly
naturally,
 though to get the precise experimental values of the mass ratios requires the  more
 careful fixing of the parameters of Table 2. 
 
 The above argument is concerned
 with the quark mass ratios. The overall scale of the quark masses, e.g. the value of $m_t$,
 is governed by the structure of the Calabi-Yau manifold. Thus from (46) one has
 \begin{equation} m_{t}={1\over{8{\sqrt 2}}}{1\over {R^3 V^{3/2}}}{1\over {\sqrt
G_{H_2}}}{\mu_3^{(u)}\over\lambda^{(u)}_3}{v\over\sqrt{2}}sin\beta.
\label{eq51}\end{equation} We had  above  that $1/\lambda^{(u)}_3=2.9\times10^4$ and detailed
calculations including the other factors in Eq.(\ref{eq46}) (and using $v$=246 GeV, $m_t$=175
GeV) yields
\begin{equation} R^3 V^{3/2}\sqrt G_{H_2}\simeq 2.1\times 10^3
\label{eq52}\end{equation} If we write $V=r^6$, where $r$ is the mean radius of the 
Calabi-Yau manifold divided by the co-ordinate radius, then for $R=1=G_{H_2}$ one finds $r\simeq 2.3$
(and smaller values of $r$ are possible if $R,\sqrt G_{H_2}>1$). While this result is
reasonable, we see that the value of $m_t$ is sensitive to the value of $r$, i.e. to the
detailed structure of the Calabi-Yau manifold.

The CKM matrix elements are more complicated functions of the model parameters, as one must
diagonalize $\lambda^{(u,d)}$ of Eqs. (\ref{eq38}, \ref{eq39}) at the electroweak scale
by making unitary transformations on $u'_L,\, u'_R,\, d'_L,\, d'_R$. (Recall that
$\lambda^{(d)}$ is relatively complicated as it is not symmetric.) The CKM matrix is then
given by the usual expression $ V_{\rm CKM}=V_{u_L}^{\dag}V_{d_L}$. The fact that the
Wolfenstein hierarchy of the CKM  matrix element is generated in Table 3 (without introducing
the Wolfenstein parameter
$\lambda$) is thus a non trivial success of the model. Currently, $V_{us}$ and
$V_{cb}$ are the best determined off diagonal elements. The model will be subjected to more
detailed test as the other CKM elements and the phases $\beta$ and $\gamma$ become better known.
 
\section{Torus Fibered Calabi-Yau Manifold with $\beta^{(0)}=0$}

In Sec 3. we had constructed a phenomenologically viable set of Yukawa textures based on the
general structure of $Z_{IJ}$ and $G_{IJ}$ arising in M-theory. In order that these quantities
have the right general size, we  assumed that the instanton charges, $\beta_i^{(0)}$, on the
physical orbifold plane at $\rho=0$ vanish and that $G_{IJ}$ for $I$ and/or $J$ in the third
generation be small (or vanish) for the u-quark sector. (More precisely, we  assumed for
simplicity that the
$SU(5)$ symmetry is still maintained for the Yukawa sector and so this condition was to hold for the
$u_L$, $d_L$ and $u_R$ quarks.) In general, it has been shown that for smooth elliptically
fibered Calabi-Yau three-folds, $\beta_i^{(0)}$ (or
$\beta_i^{(N+1)}$) can not vanish\cite{lo}. However, we will exhibit here a three generation
example of a torus fibered Calabi-Yau manifold with structure group $G=SU(5)$, with $\beta_i^{(0)}=0$,
utilizing the del Pezzo base $dP_7$. It is interesting that the additional freedom of torus
fibering (needed to generate a manifold
$Z$ with fundamental group $\pi_1(Z)=Z_2$ to break $H=SU(5)$ down to the Standard Model group by a Wilson line)
is also the additional frreedom allowing $\beta_i^{(0)}$ to vanish. At the end of this
section we will also discuss how the above third generation condition on $G_{IJ}$ might
possibly arise.

In\cite{dopw}, general rules were given for constructing realistic three generation torus fibered
vacua. We summarize now these rules and will state them for the case $G=SU(n)$, n=odd. We are
interested here in the case n=5. However it is intersting to see how this choice helps in
constructing the vacuuum we want.
\\\\
(1) We start with an elliptically fibered Calabi-Yau three fold $X$ with smooth base B
allowing a freely acting involution $\tau_X$. The torus fibered manifold,
$Z=X/\tau_X$ has two sections, $\sigma$ and $\zeta$, where
$\tau_X(\sigma)=\zeta$ and $\tau_X(\zeta)=\sigma$, and hence has $\pi_1(Z)=Z_2$. The manifold
$X$ allows the base B to be one of the del Pezzo two folds\footnote{Other possible bases are
Hirzeburch, blown up Hirzeburch or Enrique surfaces. We will not consider these here.}
$dP_r$, $r=1...8$. (For our purposes we will want $r$=7 to obtain a three generation model.)
\\\\
(2) The semi-stable holomorphic vector bundle over $X$ is given in terms of spectral data
which is specified by an effective class\footnote{Briefly, a class of holomorphic curves or linear combination of
curves with positive integer coefficients.} $\eta$ in $B$ and numbers $\lambda$ and $\kappa_i$. These
quantities obey
\begin{equation}\eta\,\,{\rm is\,\, an\,\, effective\,\, divisor\,\, class}
\label{eq53}\end{equation}and \begin{equation}\lambda-{1\over 2}\in{{\bf{Z}}};
\,\kappa_i-{1\over 2}m\in{\bf{Z}};\, m={\rm integer}.
\label{eq54}\end{equation} 
\\\\
(3) In order that the vector bundle $V$ of $X$ descend to a vector
bundle $V_Z$ of $Z$, it is required that
\begin{equation}\tau_B(\eta)=\eta
\label{eq55}\end{equation}\begin{equation}\sum\kappa_i=\eta\cdot c_1;\, i=1,1,...4\eta\cdot c_1
\label{eq56}\end{equation} (where ``$\cdot$" means intersection numbers and all unspecified Chern
classes $c_i$ are of the base $B$.) 
\\\\
(4) Three generation condition. The number of generations is
given by the third Chern class for the vector bundle on the physical orbifold plane:
$N_{\rm gen}={1\over2}c_3(V_{Z1})$. The third Chern class has been evaluated to be
\cite{gc} $c_3(V_X)=2\lambda\eta(\eta-nc_1)$, and since $X$ is a double cover of
$Z=X/\tau_X$ one has $c_3(V_{Z1})=\lambda\eta(\eta-nc_1)$. Hence the condition that there are
three generations is 
\begin{equation}\lambda\eta(\eta-nc_1)=2N_{\rm gen}=6
\label{eq57}\end{equation} (5)We assume here that the vector bundle on the distant orbifold plane
is trivial i.e. $c_2(V_2)=0$. The anomaly cancellation condition, Eq.(\ref{eq9}), then
reduces to  
\begin{equation}[W]=c_2(TX)-c_2(V_1)
\label{eq58}\end{equation} There is a class of elliptic fibers F and a new class N arising
from the blowing up of singular points on the base. The second Chern classes have been
evaluated to be \cite{dopw}
\begin{equation}c_2(TX)=12\sigma\cdot\pi^*c_1+(c_2+11 c_1^2)(F-N)+(c_2-c_1^2)N
\label{eq59}\end{equation} and 
\begin{eqnarray}\label{eq60}c_2(V)&=&\sigma\cdot\pi^*\eta-\{{1\over
{24}}(n^3-n)c_1^2-{1\over2}(\lambda^2-{1\over 4})n\eta(\eta-nc_1)-k^2\}
(F-N)\\\nonumber&-&\{{1\over {24}}(n^3-n)c_1^2-{1\over2}(\lambda^2-{1\over
4})n\eta(\eta-nc_1)-k^2+\sum\kappa_i\})N;\,\,k^2=\sum\kappa^2_i
\nonumber\end{eqnarray} (where $\pi$ is the holomorphic map $X\rightarrow B$ with
fibers
$\pi^{-1}(b)$ for point b of B and $\sigma$ is the section of $X$). Eq.(\ref{eq58}) thus
becomes 
\begin{equation} [W]=\sigma_*\omega+c(F-N)+dN\label{eq61}\end{equation} where
\begin{equation} c=c_2+[{1\over {24}}(n^3-n)+11]c_1^2-{1\over2}(\lambda^2-{1\over
4})n\eta(\eta-nc_1)-k^2\label{eq62}\end{equation}
\begin{equation} d=c_2+[{1\over {24}}(n^3-n)-1]c_1^2-{1\over2}(\lambda^2-{1\over
4})n\eta(\eta-nc_1)-k^2+\sum\kappa_i\label{eq63}\end{equation} and\begin{equation}\omega=12
c_1-\eta\label{eq64}\end{equation} Since $[W_Z]$ must represent physical holomorphic curves,
it must be an effective class. Hence 
\begin{equation}
\omega=12c_1-\eta {\rm \,\,is\,\, effective}\label{eq65}\end{equation} and\begin{equation}
c\geq0,\,d\geq0\label{eq66}\end{equation} \\\\
(6) Finally, in order that $V$ does not have a
structure group smaller than
$SU(5)$, there is a ``stability condition" \cite{gr,dopw}
\begin{equation}
\eta\geq5c_1\label{eq67}\end{equation} The above represents a set of sufficient conditions to
achieve physically viable three generation models. In addition we wish to impose the
condition that
$\beta^{(0)}_i=0$. From Eq.(\ref{eq4}) and Eq.(\ref{eq10}), this implies (on the physical
orbifold plane)
\begin{equation} \Omega\equiv c_2(V_1)-{1\over 2}c_2(TX)=0\label{eq68}\end{equation}  and thus from
(\ref{eq59}) and (\ref{eq60})
\begin{equation}\sigma.\pi^*(6 c_1-\eta)+{\tilde c}(F-N)+{\tilde d}N=0
\label{eq69}\end{equation} where
\begin{equation}{\tilde c}=c-{1\over 2}c_2-{11\over 2}c_1^2
\label{eq70}\end{equation}
\begin{equation}{\tilde d}=d-{1\over 2}c_2+{1\over 2}c_1^2
\label{eq71}\end{equation} We wish now to look for a manifold obeying the conditions of Eqs. 
(\ref{eq53}), (\ref{eq54}), (\ref{eq55}), (\ref{eq56}), (\ref{eq57}), (\ref{eq65}), 
(\ref{eq66}), (\ref{eq67}) and (\ref{eq69}), within the framework of del Pezzo bases. Some
properties of del Pezzo bases are given in\cite{dlow2,fr}. We are interested in the Chern classes
$c^2_1$ and $c_2$ which are
\begin{equation}c_1^2(dP_r)=9-r;\,c_2(dP_r)=3+r;
\label{eq72}\end{equation} and hence for $dP_7$
\begin{equation}c_1^2(dP_7)=2;\,c_2(dP_7)=10.
\label{eq73}\end{equation} We start by implementing (\ref{eq69}) by requiring
\begin{equation}\eta=6c_1
\label{eq74}\end{equation} which satisfies Eqs.(\ref{eq53}) and (\ref{eq55}). Eq.
(\ref{eq56}) then becomes
\begin{equation}\sum \kappa_i=6 c_1^2=12;\, i=1...48.
\label{eq75}\end{equation} The three generation condition Eq.(\ref{eq57}) becomes
\begin{equation}\lambda6(6-n)c_1^2=6
\label{eq76}\end{equation} and for n=5 then \begin{equation}\lambda={1\over 2}
\label{eq77}\end{equation} satisfying Eq.(\ref{eq54}). Inserting these results, 
Eqs.(\ref{eq62}) and (\ref{eq63}) become
\begin{equation}c=42-k^2;\,d=30-k^2
\label{eq78}\end{equation} and hence Eq.(\ref{eq66}) is satisfied provided 
\begin{equation}k^2\leq 30
\label{eq79}\end{equation} Further Eq.(\ref{eq74}) imples $\omega=6 c_1$ satisfying
Eq.(\ref{eq65}) and also the stability condition Eq.(\ref{eq67}). Using Eqs.(\ref{eq73}) and
(\ref{eq78}), $\tilde c$ and $\tilde d$ become 
\begin{equation}\tilde c=\tilde d=26-k^2
\label{eq80}\end{equation} Thus we can satisfy Eq.(\ref{eq68}) if we choose 
\begin{equation}k^2=\sum\kappa_i^2=26
\label{eq81}\end{equation} which clearly is consistent with Eq.(\ref{eq79}). One needs
therefore a set of $\kappa_i$ which simultaneously satisfy Eqs.(\ref{eq75}) and
(\ref{eq81}) for
$\kappa_i$ obeying Eq.(\ref{eq54}). An example of such $\kappa_i$ is 
\begin{equation}\kappa_1=\kappa_2=\kappa_3=\kappa_4=1;\,\kappa_5=2;\,\kappa_6=\kappa_7=3.
\label{eq82}\end{equation} (the remaining $\kappa_i$ being zero). This completes the
derivation that  there can exist a torus fibered Calabi-Yau with  del Pezzo base $dP_7$ for a
three generation model with fundamental group $\pi_1(Z)=Z_2$ and vanishing instanton charge
$\beta_i^{(0)}=0$ on the physical orbifold plane \footnote{There is actually one additional
point needing to verification, that the curves arising from the vanishing of the Weierstrass
discriminant can be chosen to avoid all fixed points in the base \cite{dopw}. We have not
verified this but believe that there should be sufficient freedom to achieve this.}.

We turn now to second requirement of our phenomenological fit of the Yukawa couplings that
$G_{33}$ and $G_{32}=\bar G_{23}$ be zero, or at least very small (i.e.$\stackrel{<}{\sim}
O(10^{-3})$) for $Z^u$. We have no derivation that this can be achieved, and give here only
some reasonable arguments. From the definition of $G_{IJ}$ in Eq.(\ref{eq14}), it is of course
possible that $G_{23}$ vanishes since $u_{2a}$ may be orthogonal to $u_{3\bar b}$ under the
metric
$g^{a\bar b}$. However, since $G_{IJ}$ is a hermitian matrix with positive eigenvalues, $G_{33}$ can not be precisely
zero. Thus going to a basis that diagonalizes $g^{a\bar b}$ at each point $z^a$, one would
have
\begin{equation}G_{33}={1\over vV}\int_X{\sqrt g}u_{3ax}(z)\lambda^a(z)u^x_{3\bar a}(z)
\label{eq83}\end{equation} where $\lambda^a(z)$ are the eigenvalues of $g^{a\bar b}$. 
Thus $G_{33}$ might be small if $u_{3a}$ peaked in a region of the
Calabi-Yau space different from where the eigenvalues $\lambda^a(z)$ were large for the
u-quark functions\footnote{Note that this does not imply that the $O(\epsilon)$ terms would
be similarly suppressed since these depend in $\Gamma^i_{IJ}$ on
$\partial G_{IJ}/\partial a_i$ and the full set of (1,1) harmonic forms $\omega_{ia\bar b}$
 would now enter in the integral. While one could diagonalize $g^{a\bar b}$ in
Eq.(\ref{eq83}), one can not simultaneously diagonalize the $h^{1,1}$ functions
$\omega_{ia\bar b}$ and so, a priori there is no reason for $\Gamma^i_{IJ}$ to be small.}. 
We have no explicit model that can achieve this suppression of $G_{33}$ 
(one can not explicitly
calculate $g^{a\bar b}$), but view it as a
phenomenological condition to be imposed on the Calabi-Yau manifold.

\section{Conclusion} The 11 dimensional Horava-Witten M-theory offers a fundamental framework
for the  construction of phenomenologically viable models. It allows for the possibility of
the conventional grand unified groups SU(5) or SO(10) and incorporates the unification of
gauge coupling constants at
$M_G\stackrel{\sim}{=}3\times 10^{16}$ GeV, an experimental result that has been verified to
high precision. If the elliptically fibered Calabi-Yau manifolds are generalized to torus
fibering (with e.g. two sections) three generations models can be constructed with SU(5)
symmetry broken at $M_G$ to the Standard Model group $SU(3)\times SU(2)\times U(1)$ by a
Wilson line, while still maintaining the gauge coupling unification. Under these circumstances
(unlike in standard supersymmetry grand unification) the SU(5) or SO(10) symmetries of the
Yukawa couplings (which are in fact difficult to satisfy phenomenologically) no longer need
to hold after Wilson line breaking, and the Yukawa couplings need only to obey the
$SU(3)\times SU(2)\times U(1)$ symmetry. Thus the $SU(5)$ prediction of $b-\tau$ Yukawa
unification at $M_G$ (or the SO(10) $t-b-\tau$ Yukawa unification) need not hold. Similarly,
the predictions of proton decay in standard grand unification arise from the fact tha the
superheavy color Higgs triplet and $SU(2)$ Higgs doublet lie, e.g. for $SU(5)$, 
in the $5$ and
$\bar 5$ representations and hence have common Yukawa couplings. Both of these predictions
are difficult if not, impossible to satisfy phenomenologically with minimal grand unification
models, and thus it is of interest that they need no longer be valid in M-theory with Wilson
line breaking. On the other hand, the unification of gauge coupling constants at $M_G$
is required in M-theory with Wilson line breaking, and this relation does appear to be
experimentally valid. 

The form of the Yukawa couplings and how to accommodate the the remarkable hierarchies of the
quark lepton mass ratios and the CKM matrix elements, is one of the major unresolved problems
in the Standard Model. The conventional approach is to assume a Yukawa texture at $M_G$ where
high powers of the Wolfenstein parameter $\lambda\stackrel{\sim}{=}0.2$ appear to obtain large
third generation masses and small off diagonal elements at the electroweak scale. (See e.g.
Table 1.) M theory offers a unique alternate possibility where the hierarchies are encoded in
the Kahler metric $Z_{IJ}$, which must be diagonalized and rescaled to the unit matrix to
obtain the Yukawa interactions of the canonically normalized fields. Unlike supergravity
grand unification, where the Kahler potential is undetermined, the general structure of the
Kahler potential is determined in M-theory. In particular M-theory offers the possibility
that if the 5 branes required by the non-standard embeddings cluster near the distant
orbifold plane, then a parameter of smallness
$d=1-z_n\stackrel{\sim}{=}0.1$ can naturally enter in the third generation contributions to
$Z_{IJ}$, leaving other entries $O(1)$. A model based on this possibility, in accord with all
quark mass and CKM data, was given in Sec. 3. (See Tables 2-4.) No high powers of $d$ or
exceptional fine tuning of parameters were needed to generate the experimental hierarchies.

In order to obtain the desired form of the Kahler metric, it was reasonable to require that
the instanton charges, $\beta_i^{(0)}$, on the physical orbifold plane vanish. (This is not
possible for elliptically fibered manifolds\cite{lo}). In Sec.4, we constructed an example
of a three generation torus fibered  Calabi-Yau manifold with SU(5) structure group and del
Pezzo base $dP_7$ where the $\beta_i^{(0)}$ vanish. (Lower del Pezzo bases would give rise to
models with more generations.) The fact that this can be done is somewhat remarkable: one
must find consistent solutions to certain Diophantine equations. Why solutions exist at all is
unclear. It is interesting to note, however, that the generalization from elliptic to torus
fibration, needed to allow for Wilson line breaking to the Standard Model gauge group, is
also what is needed to achieve $\beta_i^{(0)}=0$.

If we assume that the vacuum discussed above is valid, one may speculate as to what may be
expected on the distant orbifold plane. We first consider the gauge kinetic functions of
Eq.(\ref{eq19}). Since $\beta_i^{(0)}=0$, and if we assume $Im Z_n$ is small (since
experimentally the electric dipole moment of the neutron is very small if not zero) then 
\begin{equation} f^{(1)}\stackrel{\sim}{=}S+\epsilon T^i\sum^N_{n=1}(1-z_n)^2\beta^{(n)}_i
\label{eq84}\end{equation} i.e the gauge function on the physical orbifold plane deviates by
only a small amount from the dilaton S when $d_n=1-z_n\simeq 0.1$ (as suugested by the Yukawa
coupling analysis). On the other hand, the gauge kinetic function $f^{(2)}$ on the hidden plane would
appear a priori to have a large correction. However, the magnetic charges on the 5-branes
obey a constraint 
\begin{equation}
\sum^N_{n=1}\beta^{(n)}_i+\beta^{(N+1)}_i=0
\label{eq85}\end{equation} since $\beta_i^{(0)}$=0. Hence to first order one has 
\begin{equation}   f^{(2)}\stackrel{\sim}{=}S-2\epsilon T^i\sum^N_{n=1}d_n \beta^{(n)}_i
\label{eq86}\end{equation} and while this correction is larger than $f^{(1)}$, it may still be
sufficiently small to validate the perturbation expansion in $\epsilon$ since
$d_n$ is small. A similar phenomena occurs for the Kahler metric. The expression on the
distant orbifold plane is obtained by replacing $z_n$ by
$1-z_n$ in Eq.(\ref{eq21}) yielding
\begin{equation}   Z^{(2)}_{IJ}\stackrel{\sim}{=}e^{-K_T/3}[G_{IJ}+{\epsilon\over V}\tilde
\Gamma^i_{IJ}\sum^N_{n=1}d_n \beta^{(n)}_i]
\label{eq87}\end{equation} The corrections from the $O(\epsilon)$ terms are now about 10
times larger than on the physical orbifold plane, suggesting a reduction of hierarchy of the
Yukawa mass scales on the second orbifold plane. We have here assumed that the full $E_8$ symmetry holds on the hidden
plane. However, more complicated models where the structure group on the hidden plane is
non-trivial might also be considered.

As in other string related models, M-theory still suffers from a lack of understanding of
some of the basic principles of the model, e.g. how does supersymmetry break, what picks out
the correct (physical) vacuum etc. This is reflected in one of the weaknesses of the model of
Yukawa couplings we have discussed, i.e. while all the mass ratios can naturally be obtained,
the overall scale of the quark masses, e.g. the value of $m_t$, is sensitive to the volume
modulus $V$ of the Calabi-Yau manifold. This appears to be a generic problem for this class
of models, and perhaps the experimental value of $m_t$ can give further phenomenological
insight into the structure of the correct vacuum state.

Finally we mention that we have not discussed here the lepton Yukawa couplings. But the
charged leptons can be treated with textures similar to those in the quark sector. It is
also
possible to use these textures to investigate the structure of the supersymmetry soft
breaking masses.

\section{Acknoledgements} We should like to thank B. Ovrut for a  
useful conversation and A. Lukas for a helpful clarification. This
work was supported in part by the National Science Foundation Grant PHY-9722090.

\end{document}